\documentclass[12pt]{article}
\newlength{\vshift}
\input{epsf.tex}
\newlength{\hshift}

\usepackage{amssymb}
\usepackage{amsmath}
\usepackage{amsthm}
\usepackage{amsopn}
\def\beq{\begin{equation}}
\def\eeq{\end{equation}}
\def\bea{\begin{eqnarray}}
\def\eea{\end{eqnarray}}
\def\no{\nonumber}
\def\y{y\!\!\!/}

\begin{document}
\vspace*{3cm}
\begin{center}
{\Large\bf{Nucleon-Nucleon Scattering in a Strong External
Magnetic Field and the Neutrino Emissivity}}\vskip 4em{{\bf E. Bavarsad$^{a}$}, {\bf M. Haghighat$^{a,b}$} and
{\bf R. Mohammadi$^{a}$}}\vskip 1em
{\it a) Department of Physics, Isfahan University of Technology,\\ Isfahan 84156-83111, Iran\\}
{\it b) Department of Physics, University of Alabama,\\ Tuscaloosa,
Alabama 35487, USA\\}
\end{center}
\vspace*{1.5cm}
\begin{abstract}
The nucleon-nucleon scattering in a large magnetic background is
considered to find its potential to change the neutrino emissivity
of the neutron stars. For this purpose we consider the one-pion-exchange
approximation to find the NN cross-section in a
background field as large as
$10^{15}\texttt{G}-10^{18}\texttt{G}$. We show that the NN
cross-section in  neutron stars with temperatures in the range
0.1-5 \texttt{MeV} can be changed up to the one order of magnitude
with respect to the one in the absence of the magnetic field. In
the limit of the soft neutrino emission the neutrino emissivity
can be written in terms of the NN scattering amplitude therefore
the large magnetic fields can dramatically change the neutrino
emissivity of the neutron stars as well.
\end{abstract}
\newpage
\section{Introduction}
Neutron stars in general can have magnetic fields as large as
$10^{12}\texttt{G}-10^{18}\texttt{G}$ \cite{haberl},
\cite{tsuruta}, \cite{thompson}, \cite{blandford}, \cite{Duncan},
\cite{Mereghetti} while the timing observations indicate that the
surface magnetic fields of the radio pulsars are about
$10^{12}\texttt{G}$ \cite{manchester}, by equating magnetic field
energy to the gravitational binding energy, one can estimate an
upper limit for strength of the magnetic field to be
$B\sim10^{18}\texttt{G}$ \cite{max}. Meanwhile cooling processes
in the neutron stars formation that has long been considered by
many physicists, can be affected by such a large magnetic field in
two ways. It can change  the electron or the photon conductivities
\cite{itoh},\cite{tsuruta}, \cite{aguilera} as well as the wave
functions and the propagators of the particles involving in the
cooling processes. One of the most important reactions during the
thermal evolution of the neutron stars is the emission of neutrino
via the bremsstrahlung process
\begin{equation}\label{01}
NN\rightarrow NN+\nu\bar{\nu}.
\end{equation}
In the first sight it seems that since the mass of the nucleons
are about 1\texttt{Gev} therefore the magnetic fields even as
large as $10^{18}\texttt{G}$ can not significantly change the
cross section. However one should note that there are more energy
scales besides the mass of the nucleons. In fact one of the scales
is the temperature of the evolving neutron star which becomes less
than 1\texttt{MeV} in a few seconds of the late-time cooling phase
and is comparable with $B\sim10^{13}\texttt{G}$. Furthermore if
one consider one-pion-exchange (OPE) for the nucleon-nucleon
interaction then the magnetic fields of the order of
$10^{17}\texttt{G}$ and more can change the strong coupling
constant as well. Meanwhile in such a low temperature era the
bremsstrahlung process can only produce the soft neutrinos with
respect to the initial energies and therefore the neutrino
emissivity can be related to the on-shell NN scattering amplitude
\cite{hanhart}. Thus to find the neutrino emissivity one can study
the nucleon-nucleon scattering separately. The amplitude can be
even constructed from NN scattering data, as is done in
\cite{hanhart} as a model independent way to describe the neutrino
emissivity. But the usual studies on  the nucleon-nucleon
scattering in OPE approximation \cite{friman} and more precisely
the existing data are in the absence of the magnetic field. Our
aim in this paper is to examine the magnetic field dependence of
the nucleon-nucleon scattering in the OPE approximation when soft
neutrino approximation is applicable.
\par The paper is organized as follows:
in section 2 we give the field operators of the proton and
neutrons in an external magnetic field. Consequently  we obtain
the cross section of the scattering in the magnetic field in
section 3. An estimation for the effect of strong magnetic field on the neutrino
emissivity is presented in section 4.
In section 5 we give some concluding remarks. Appendix
A is devoted to solving the Dirac equation for proton and neutron
in an external magnetic field. In Appendix B we present a
simplified expression for the amplitude matrix elements of the
scattering process.
\section{Fermion field operator}
To study the proton-neutron scattering  in a
strong external magnetic field, the field operators of the proton
and neutron are needed. Since the field operators contain wave
functions and energy spectrum, Dirac equation in presence of an
external magnetic field must be solved. We choose the external
magnetic field to be in the positive $z$-axis $\vec{B}=B\hat{k}$.
Because of gauge invariance there is a freedom in the choice of
vector potential $A_{\mu}(x)$, thus for simplicity we choose
\begin{equation}\label{101}
A_{\mu}(x)=(0,yB,0,0).
\end{equation}
This choice of vector potential pushes all the  $y$-dependence into  the spinors alone and not in the  phases, so it is
useful to introduce the following notations
\begin{eqnarray}\label{102}
x^{\mu}_{\y}&=&(t,x,0,z),\no\\
\vec{V}_{\y}&=&(V_{x},V_{z}),
\end{eqnarray}
where $\vec{V}_{\y}$ is any 2-vector. The solutions of the Dirac
equation for proton and neutron in the external magnetic field
(\ref{101}) are presented in the Appendix A. By using this
solutions the proton field operator can be written as
\begin{equation}\label{103}
\psi_{p}(x)=\sum_{n=0}^{\infty}\sum_{s=\pm1}\int\frac{d^{2}\vec{p}_{\y}}{(2\pi)^{2}}N_{s}
[a_{s}(n,\vec{p}_{\y})U_{s}(y,n,\vec{p}_{\y})e^{-ip.x_{\y}}+
b_{s}^{\dag}(n,\vec{p}_{\y})V_{s}(y,n,\vec{p}_{\y})e^{+ip.x_{\y}}],
\end{equation}
where the creation and annihilation operators obey the following
anticommutation relations
\begin{equation}\label{104}
\{a_{s}(n,\vec{p}_{\y}),a_{s'}^{\dag}(n',\vec{p}'_{\y})\}=
\{b_{s}(n,\vec{p}_{\y}),b_{s'}^{\dag}(n',\vec{p}'_{\y})\}=
(2\pi)^{2}\delta_{s,s'}\delta_{n,n'}\delta^{2}_{\y}(\vec{p}'-\vec{p}).
\end{equation}
With the help of (\ref{104}), the explicit form of the
wave functions given in the equations (\ref{a16}), (\ref{a19}) and the
completeness relation (\ref{a8}), it can be shown that the  equal-time
anticommutation relation is satisfied
\begin{equation}\label{105}
\{\psi_{p}(t,\vec{x}),\psi_{p}^{\dag}(t,\vec{x}')\}=
\delta^{3}(\vec{x}-\vec{x}').
\end{equation}
For convenience we  normalize our 1-particle states in a
box with dimensions $L_{x}L_{y}L_{z}=V$ such that the proton state
is
\begin{equation}\label{106}
\mid p(n,\vec{p}_{\y},s)>=\frac{1}{\sqrt{L_{x}L_{z}}}
a_{s}^{\dag}(n,\vec{p}_{\y})\mid 0>.
\end{equation}
Similarly for the neutron field operator one has
\begin{equation}\label{107}
\psi_{n}(x)=\sum_{s=\pm1}\int\frac{d^{3}\vec{p}}{(2\pi)^{3}}N_{s}
[a_{s}(\vec{p})U_{s}(\vec{p})e^{-ip.x}+b_{s}^{\dag}(\vec{p})V_{s}(\vec{p})e^{+ip.x}],
\end{equation}
in which the creation and annihilation operators obey the
following anticommutation relations
\begin{equation}\label{108}
\{a_{s}(\vec{p}),a_{s'}^{\dag}(\vec{p}')\}=
\{b_{s}(\vec{p}),b_{s'}^{\dag}(\vec{p}')\}=
(2\pi)^{3}\delta_{s,s'}\delta^{3}(\vec{p}'-\vec{p}).
\end{equation}
By using the relations (\ref{108}) and the explicit form of the wave
functions given in the equations (\ref{a21}) and (\ref{a24}) it is easy to show
that equal-time anticommutation relation is satisfied
\begin{equation}\label{109}
\{\psi_{n}(t,\vec{x}),\psi_{n}^{\dag}(t,\vec{x}')\}=
\delta^{3}(\vec{x}-\vec{x}').
\end{equation}
Box normalization condition for the neutron 1-particle state is
\begin{equation}\label{110}
\mid n(\vec{p},s)>=\frac{1}{\sqrt{V}}a_{s}^{\dag}(\vec{p})\mid 0>.
\end{equation}
\section{Cross section of the scattering in the external magnetic field}
Our aim in this section is to explore the scattering cross section of
$p+n\rightarrow p+n$ in the background of a strong magnetic
field, $\sigma_{B}$, using the one-pion-exchange approximation. The deferential cross section can be
written as
\begin{equation}\label{201}
d\sigma=\frac{V}{|\vec{v}_{p}-\vec{v}_{n}|}\frac{\overline{|S|^{2}}}{T}d\rho,
\end{equation}
where $|\vec{v}_{p}-\vec{v}_{n}|/V$ is the flux of the
incident particles, $T$ is the time interval, $\overline{|S|^{2}}$
is the average of the squared scattering matrix elements and $d\rho$ is the
deferential phase space of the final states. The effective low
energy Lagrangian for pion-nucleon interaction is
\begin{equation}\label{202}
\mathcal{L}_{int}=g_{\pi N}
[\bar{\psi}_{p}\gamma^{5}\psi_{p}\phi_{\pi^{0}}+
\sqrt{2}\bar{\psi}_{n}\gamma^{5}\psi_{p}\phi_{\pi^{-}}+
\sqrt{2}\bar{\psi}_{p}\gamma^{5}\psi_{n}\phi_{\pi^{+}}-
\bar{\psi}_{n}\gamma^{5}\psi_{n}\phi_{\pi^{0}}],
\end{equation}
where $g_{\pi N}\simeq 14$ is the pion-nucleon coupling constant
and $\psi_{n}$, $\psi_{p}$ and $\phi$ are neutron,
proton and pion fields, respectively. To obtain the scattering matrix
elements in the external magnetic field we use the standard method where the wave functions and propagators are
modified with the external magnetic field while the vertices are leaved
unchanged. The proton-neutron scattering at the tree level involves two
diagrams, one is mediated by the neutral pion $S_{nc}$
\begin{eqnarray}\label{211}
S_{nc}=<p'(n',\vec{p}_{\y}',s'), n'(\vec{k}',r')|
T\{\int d^{4}x'd^{4}x\bar{\psi}_{p}(x')(ig_{\pi N}\gamma^{5})
\psi_{p}(x')\phi_{\pi^{0}}(x')\\\no
\bar{\psi}_{n}(x)(-ig_{\pi N}\gamma^{5})
\psi_{n}(x)\phi_{\pi^{0}}(x)\}|p(n,\vec{p}_{\y},s),n(\vec{k},r)>,
\end{eqnarray}
and the other is mediated by the charged pion $S_{cc}$
\begin{eqnarray}\label{212}
S_{cc}=<p'(n',\vec{p}_{\y}',s'), n'(\vec{k}',r')|
T\{\int d^{4}x'd^{4}x\bar{\psi}_{p}(x')(i\sqrt{2}g_{\pi N}\gamma^{5})
\psi_{n}(x')\phi_{\pi^{+}}(x')\\\no
\bar{\psi}_{n}(x)(i\sqrt{2}g_{\pi N}\gamma^{5})
\psi_{p}(x)\phi_{\pi^{-}}(x)\}|p(n,\vec{p}_{\y},s),n(\vec{k},r)>,
\end{eqnarray}
where $p$ ($n$) and $p'$ ($n'$), respectively, correspond to the incoming and outgoing
protons (neutrons). Since the neutral pion has no charge its
propagator in the external magnetic field does not change, but
the propagator of charged pion will be modified. In fact the propagator of the charged
pion as a charged scalar particle in an external magnetic field can be obtained as
\cite{Erdas}
\begin{eqnarray}\label{214}
<T\{\phi_{\pi^{+}}(x')\phi_{\pi^{-}}(x)\}>=
\exp(-ie\int_{x}^{x'}d\xi_{\mu}\{A^{\mu}+\frac{1}{2}F^{\mu\nu}(\xi-x)_{\nu}\})
\int\frac{d^{4}q}{(2\pi)^{4}}e^{-iq.(x'-x)}\no\\
\times\int_{0}^{\infty}\frac{{d s}}{\cos(eBs)}
\exp(is\{q_{\|}^{2}-\frac{\tan(eBs)}{eBs}q_{\bot}^{2}-m_{\pi}^{2}\}),
\end{eqnarray}
in which $F_{\mu\nu}$ is the field strength tensor of the external
magnetic field and
\begin{equation}
q_{\|}^{2}=(q_{0})^{2}-(q_{3})^{2},
~~~~~~~~~~q_{\bot}^{2}=(q_{1})^{2}+(q_{2})^{2}.
\end{equation}
The temperature of a neutron star in a few seconds in the late-time cooling phase, drops to less than 1 \texttt{MeV} therefore the energy is sufficiently low to use the four-fermion approximation, $q^{2}/m_{\pi}^{2}=0$. In this case the propagator (\ref{214}) can be cast into
\begin{eqnarray}\label{215}
<T\{\phi_{\pi^{+}}(x')\phi_{\pi^{-}}(x)\}>&=&\frac{-i}{m_{\pi}^{2}}
D(eB/m_{\pi}^{2})\delta^{4}(x'-x);\no\\
D(eB/m_{\pi}^{2})&=&\int_{0}^{\infty}ds
\frac{e^{-s}}{\cosh(\frac{eBs}{m_{\pi}^{2}})}.
\end{eqnarray}
\begin{figure}
\centerline{\epsfysize=3in\epsfxsize=3in\epsffile{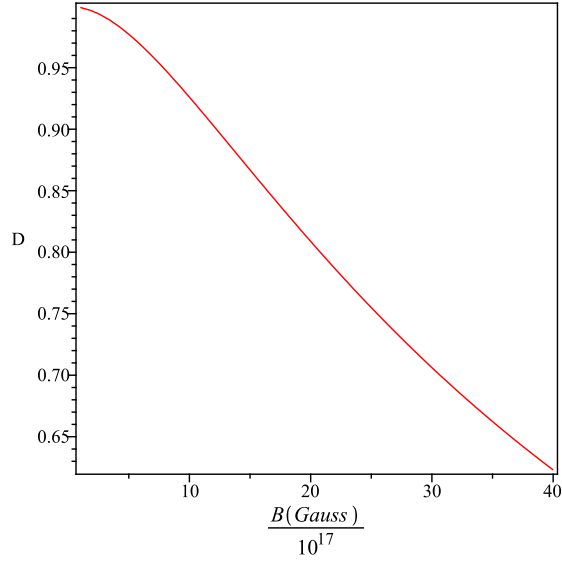}}\caption{The B-dependence of the squared coupling
constant. The variation starts from $10^{15}\texttt{G}$ but the
value decreases significantly from $10^{17}\texttt{G}$.}
\label{figA_1}
\end{figure}
In the four-fermion approximation, $D$ actually shows how the
strong coupling constant depends on the external magnetic field.
The B-dependence of the coupling constant is explicitly shown in
Fig. 1. By using the explicit form of the proton and neutron field
operators (\ref{103}) and (\ref{107}) and the definitions of the
1-particle states (\ref{106}) and (\ref{110}), the scattering
matrix elements can be obtained as follows
\begin{equation}\label{216}
S_{nc(cc)}=\frac{1}{L_{x}L_{z}V}N_{p}N_{n}N_{p'}N_{n'}
(2\pi)^{3}\delta_{\y}^{3}(p'+k'-p-k) (-i\mathcal{M}_{nc(cc)}),
\end{equation}
where the neutral current scattering amplitude is given by
\begin{eqnarray}\label{217}
\mathcal{M}_{nc}=\frac{g_{\pi N}^{2}}{m_{\pi}^{2}}\int{dy}
e^{-i(k'_{y}-k_{y})y}[\bar{U}_{p'}(y,n',\vec{p}_{\y}',s')\gamma^{5}U_{p}(y,n,\vec{p}_{\y},s)]
[\bar{U}_{n'}(\vec{k}',r')\gamma^{5}U_{n}(\vec{k},r)],
\end{eqnarray}
and the charged current is
\begin{eqnarray}\label{218}
\mathcal{M}_{cc}=2D\frac{g_{\pi N}^{2}}{m_{\pi}^{2}}\int{dy}
e^{-i(k'_{y}-k_{y})y}[\bar{U}_{p'}(y,n',\vec{p}_{\y}',s')\gamma^{5}U_{n}(\vec{k},r)]
[\bar{U}_{n'}(\vec{k}',r')\gamma^{5}U_{p}(y,n,\vec{p}_{\y},s)].
\end{eqnarray}
Now we are ready to calculate the scattering cross
section in the external magnetic field. For this purpose we have
\begin{equation}\label{221}
\overline{|S|^{2}}=\frac{1}{(2s_{p}+1)}
\frac{1}{(2s_{n}+1)}\sum_{spin}|S_{nc}+S_{cc}|^{2},
\end{equation}
where using the normalization relation for the Dirac
$\delta$-functions
\begin{equation}\label{222}
(\delta_{\y}^{3}(p'+k'-p-k))^{2}=\frac{1}{(2\pi)^{3}}
TL_{x}L_{z}\delta_{\y}^{3}(p'+k'-p-k),
\end{equation}
cast the squared amplitude into
\begin{eqnarray}\label{223}
|S|^{2}&=&\frac{TL_{x}L_{z}}{(L_{x}L_{z}V)^{2}}
N_{p}^{2}N_{n}^{2}N_{p'}^{2}N_{n'}^{2}(2\pi)^{3}
\delta_{\y}^{3}(p'+k'-p-k)|\mathcal{M}|^{2}.
\end{eqnarray}
Now the equations (\ref{201}), (\ref{223}) and the deferential phase
space of the final states in the external magnetic field
\begin{equation}\label{224}
d\rho=\frac{d^{2}p'_{\y}}{(2\pi)^{2}}L_{x}L_{z}
\frac{d^{3}k'}{(2\pi)^{3}}V,
\end{equation}
can be used to find the corresponding deferential cross section as follows
\begin{eqnarray}\label{225}
d\sigma_{B}=\frac{1}{|\vec{v}_{p}-\vec{v}_{n}|}\frac{1}{(2s_{p}+1)(2s_{n}+1)}
\sum_{spins}N_{p}^{2}N_{n}^{2}\frac{d^{2}p'_{\y}}{(2\pi)^{2}}
\frac{d^{3}k'}{(2\pi)^{3}}N_{p'}^{2}N_{n'}^{2}\no\\
\times(2\pi)^{3}\delta_{\y}^{3}(p'+k'-p-k)|\mathcal{M}|^{2}.
\end{eqnarray}
Since the nucleons in the neutron star are nonrelativistic their
energies and wave functions can be expanded up to the order of
$\mathcal{O}(1/m_{N})$ and the amplitude $\mathcal{M}$ can be
obtained analytically as is given in the Appendix B. Meanwhile
one can easily show that
\begin{equation}\label{226}
N_{p}, N_{n}, N_{p'}, N_{n'}=1+\mathcal{O}(1/m_{N}^{2}).
\end{equation}
Therefore the leading term in the scattering cross section
(\ref{225}) is
\begin{eqnarray}\label{227}
\sigma_{B}&=&\frac{1}{|\vec{v}_{p}-\vec{v}_{n}|}
\sum_{n'=0}^{\infty}\int\frac{d^{2}p'_{\y}}{(2\pi)^{2}}
\frac{d^{3}k'}{(2\pi)^{3}}\no\\
&\times&\frac{1}{(2s_{p}+1)(2s_{n}+1)}
\sum_{spins}(2\pi)^{3}
\delta_{\y}^{3}(p'+k'-p-k)|\mathcal{M}|^{2}.
\end{eqnarray}
After integrating over $p'_{x}$, $k'_{z}$ with the help of the Dirac
$\delta$-functions and defining
\begin{equation}\label{229}
k'_{x}=k'_{\bot}\cos\theta',~~~~~~~~~ k'_{y}=k'_{\bot}\sin\theta',
\end{equation}
equation (\ref{227}) leads to
\begin{eqnarray}\label{2210}
\sigma_{B}&=&\frac{1}{4\pi^{2}|\vec{v}_{p}-\vec{v}_{n}|}
\sum_{n'=0}^{\infty}\int_{-\infty}^{+\infty}
dp'_{z}\int_{0}^{2\pi}d\theta'\int_{0}^{+\infty}
dk'_{\bot}k'_{\bot}\no\\
&\times&\frac{1}{(2s_{p}+1)(2s_{n}+1)}
\sum_{spins}\delta(E_{p}'+E_{n}'-E_{p}-E_{n})
|\mathcal{M}|^{2}.
\end{eqnarray}
By using the nonrelativistic expansion of the energies as are given in the
equations (\ref{b12}) and (\ref{b13}) one has
\begin{eqnarray}\label{2211}
0&=&E_{p}'+E_{n}'-E_{p}-E_{n}\no\\&=&\frac{1}{2m_{N}}\left(k_{\bot}^{\prime2}
+2(p'_{z}-\frac{1}{2}(p_{z}+k_{z}))^{2}-2eB(n'_{max}-n')\right),
\end{eqnarray}
where $n'_{max}$ is defined as
\begin{equation}\label{2212}
n'_{max}\equiv n+\frac{K_{p}}{2}(s'-s)
+\frac{K_{n}}{2}(r'-r)+\frac{1}{2eB}(k_{\bot}^{2}+\frac{1}{2}(p_{z}-k_{z})^{2}).
\end{equation}
The positive solution of (\ref{2211}) is
\begin{equation}\label{2213}
k_{\bot0}=\sqrt{2eB(n'_{max}-n')-2(p'_{z}-\frac{1}{2}(p_{z}+k_{z}))^{2}}.
\end{equation}
Now one can perform the integration over $k'_{\bot}$ to find
\begin{equation}\label{2214}
\sigma_{B}=\frac{m_{N}}{4\pi^{2}|\vec{v}_{p}-\vec{v}_{n}|}
\frac{1}{(2s_{p}+1)(2s_{n}+1)}\sum_{spins}
\sum_{n'=0}^{\mathcal{N}'_{max}}
\int_{\mathcal{P}'_{z-}}^{\mathcal{P}'_{z+}}
dp'_{z}\int_{0}^{2\pi}d\theta'|\mathcal{M}|^{2},
\end{equation}
where
\begin{equation}\label{2215}
\mathcal{N}'_{max}=\mathrm{INT}(n'_{max}),
\end{equation}
while the integration bounds over $p'_{z}$ can be determined by the following equation
\begin{equation}\label{2216}
\mathcal{P}'_{z\pm}=\frac{1}{2}(p_{z}+k_{z})\pm\sqrt{eB(n'_{max}-n')}.
\end{equation}
To evaluate $\sigma_{B}$ in the equation (\ref{2214}) first one
needs to determine the maximum Landau-level that proton can occupy
in the initial state, i.e.  $\mathcal{N}_{max}$. Since we would like
to consider the process (\ref{01}) in the temperature range $T\sim
1\texttt{Mev}$ \cite{tsuruta}, \cite{blandford}, \cite{Hannestad}
therefore nucleons are nonrelativistic and the energy eigenvalues
for the protons and neutrons can be given by (\ref{b12}) and
(\ref{b13}), respectively. According to the Virial theorem, each
nucleon at a temperature $T$ has a momentum for each component about
$(m_{N}T)^{1/2}$. Therefore one can easily obtain the maximum
Landau-level by comparing the magnetic energy contribution to the
proton $neB/m_{N}$, see (\ref{b12}), with the thermal energy as
$\mathcal{N}_{max}=\mathrm{INT}(m_{N}T/eB)$.
\par In this section
we obtained the cross-section in the background of a strong
magnetic field $\sigma_{B}$ by considering OPE in the four-fermion
approximation. For simplicity we calculate the cross-section in
the head-on collisions i.e. $(\vec{p}+\vec{k})_{\y}=0$. Since we
would like to find if the magnetic field can significantly change
the thermal evolution of the neutron stars, it would be better to
normalize our result with the cross-section in the absence of the
magnetic field, $\sigma_{0}$, but with the same approximations
and conditions i.e. $(\vec{p}+\vec{k})=0$, or
\begin{equation}\label{31}
\sigma_{0}=\frac{2g_{\pi N}^{4}}{3\pi
m_{\pi}^{4}|\vec{v}_{p}-\vec{v}_{n}|}
\frac{|\vec{p}|^{5}}{m_{N}^{3}}.
\end{equation}
Therefore the deviation $\frac{\sigma_B}{\sigma_0}$ as a function of the
magnetic field and temperature can be given by
\begin{eqnarray}\label{32}
\Gamma(B,T)&\equiv&\frac{1}{(1+\mathcal{N}_{max})}\sum_{n=0}^{\mathcal{N}_{max}}
\frac{\sigma_{B}}{\sigma_{0}}=\frac{3}{8\pi}(\frac{m_{N}}{|\vec{p}|})^{5}\frac{1}{(1+\mathcal{N}_{max})}
\sum_{n=0}^{\mathcal{N}_{max}}\no\\&\times&\frac{1}{(2s_{p}+1)(2s_{n}+1)}\sum_{spins}\sum_{n'=0}^{\mathcal{N}'_{max}}
\int_{\mathcal{P}'_{z-}}^{\mathcal{P}'_{z+}}
\frac{dp'_{z}}{m_{N}}\int_{0}^{2\pi}d\theta'\widetilde{|\mathcal{M}|^{2}},
\end{eqnarray}
where
\begin{equation}\label{33}
\widetilde{|\mathcal{M}|^{2}}=
\frac{m_{\pi}^{4}}{g_{\pi N}^{4}}|\mathcal{M}|^{2}.
\end{equation}
The variation of $\Gamma$ as a function of the external magnetic
field for different temperatures are shown in the Figures 2 and 3.
\begin{figure}\vspace{-1cm}
\centerline{\epsfysize=4in\epsfxsize=6in\epsffile{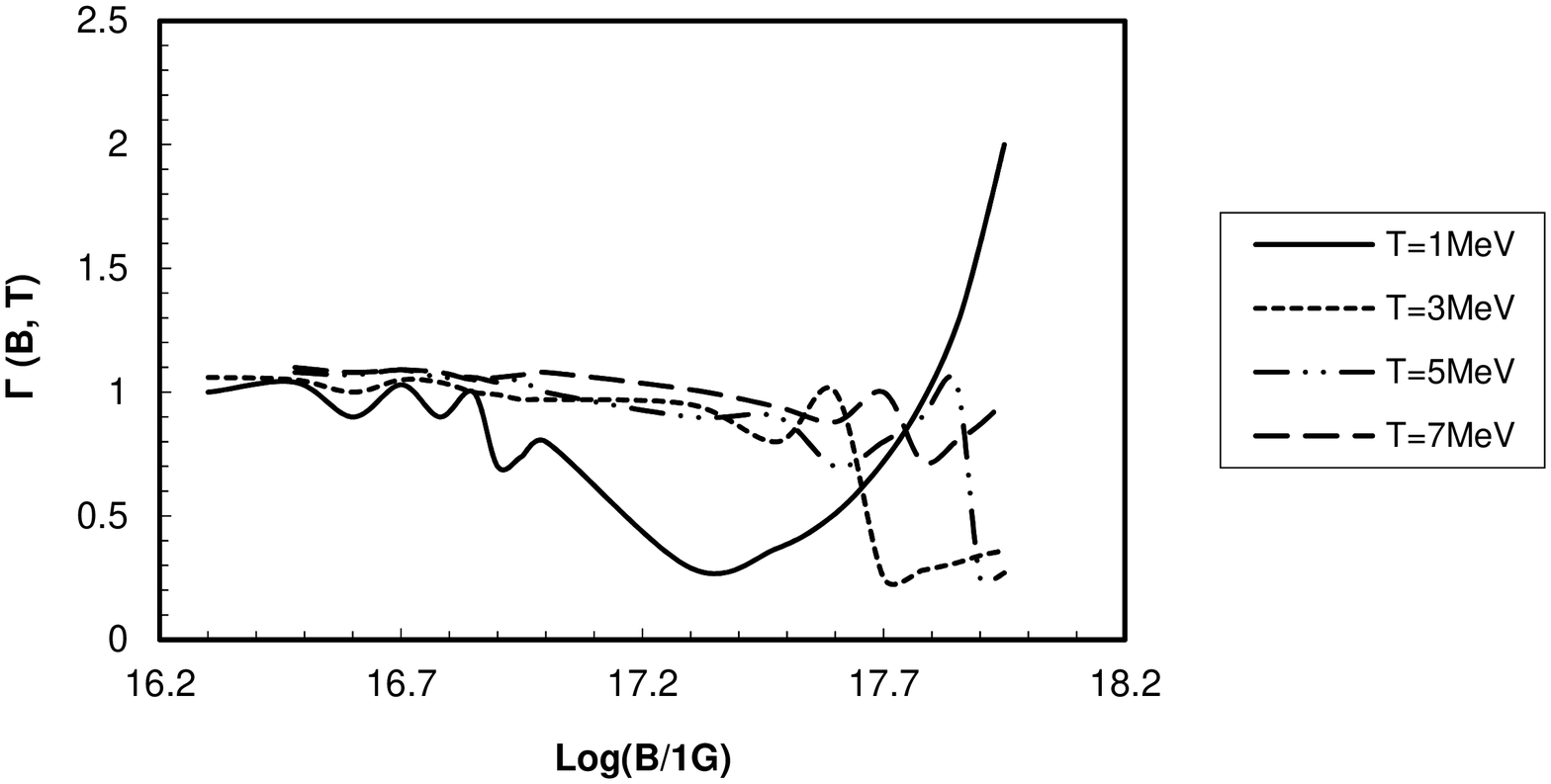}}\caption{The variation of the $\Gamma$ in terms of
$Log(B/1\texttt{G})$ for different temperatures.}\label{fig2}
\end{figure}
\begin{figure}\vspace{-2cm}
\centerline{\epsfysize=4in\epsfxsize=6in\epsffile{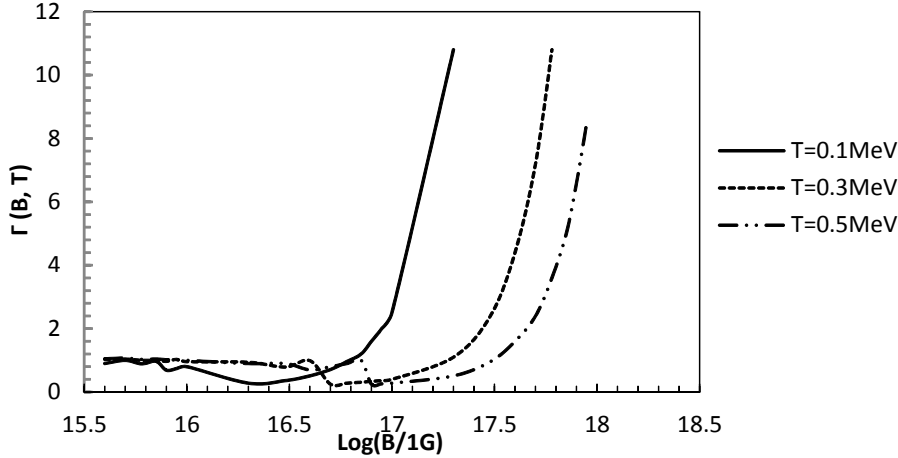}}\caption{The
variation of the $\Gamma$ in terms of $Log(B/1\texttt{G})$ for
different temperatures.} \label{fig3}
\end{figure}
\section{Neutrino emissivity}
\begin{figure}\vspace{-13cm}
\centerline{\epsfysize=8in\epsfxsize=8in\epsffile{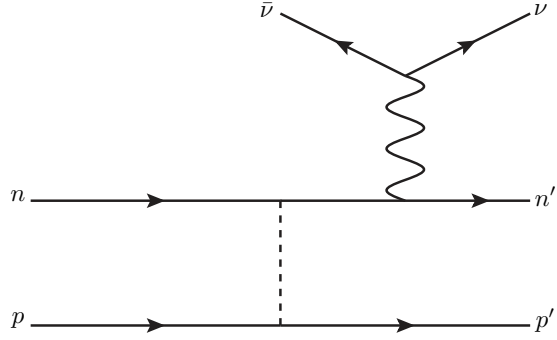}}\vspace{-1cm}
\caption{Feynman diagram for neutrino emissivity.}\label{fig4}
\end{figure}

\begin{figure}\vspace{-2cm}
\centerline{\epsfysize=4in\epsfxsize=6in\epsffile{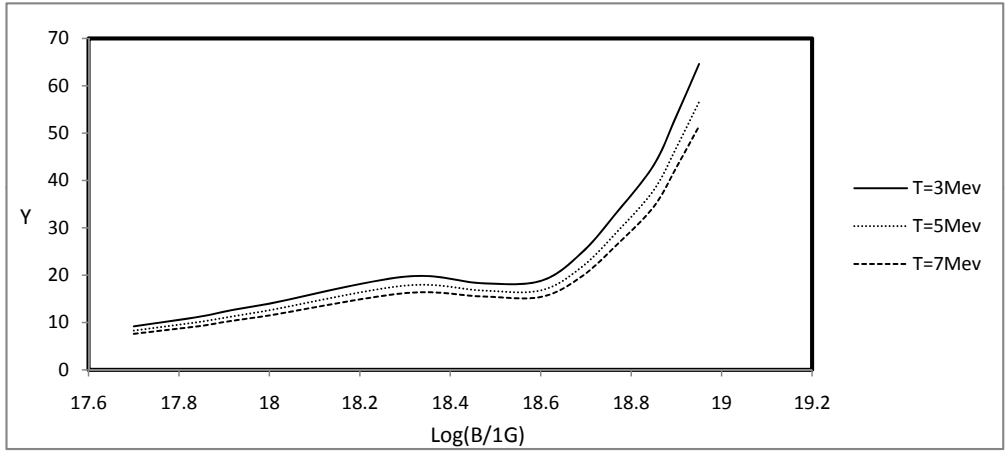}}
\caption{$\Upsilon=\frac{\epsilon_{B}}{\epsilon_{0}}$ is plotted as
function of magnetic field for different temperatures.}\label{fig5}
\end{figure}

The NN-scattering cross section in the magnetic fields as large as
$10^{15}\texttt{G}-10^{18}\texttt{G}$ shows that the strong magnetic
filed of the neutron stars might change the neutrino emissivity and
therefore the cooling process.  The rate of emission and absorbtion
of neutrino in the strong magnetic field of a neutron star have been
already considered for the Urca process in \cite{Liu},
\cite{neu-emmission} and \cite{neu-absor}.  In \cite{Liu} for a
strange stars with all charged particles in the lowest Landau level,
it is shown that the magnetic field accelerates the cooling process.
While in \cite{neu-emmission} it is found that the direct Urca
emissivity is strongly enhanced and oscillates as a function of the
magnetic field which globally tends to average out.  Here we would
like to consider the effect of strong magnetic field on the neutrino
bremsstrahlung process (\ref{01}).  To obtain the emissivity at the
tree level one should consider all Feynman diagrams at this level.
Here for simplicity we just consider the diagram given in
Fig.\ref{fig4} to give an order of magnitude for the emissivity and
leave the complete calculation for future study. For this purpose we
use
 the pion-nucleon interaction
Lagrangian (\ref{202}) and the weak interactions Lagrangian
\begin{equation}\label{401}
\mathcal{L}_{int}=\frac{g}{4\cos\theta_{\textmd{w}}}
[\bar{\psi}_{\nu}\gamma^{\mu}(1-\gamma^{5})\psi_{\nu}
-\bar{\psi}_{n}\gamma^{\mu}(1-\gamma^{5})\psi_{n}+
\bar{\psi}_{p}\gamma^{\mu}(1-4\sin^{2}\theta_{\textmd{w}}-\gamma^{5})\psi_{p}]Z^{0}_{\mu}.
\end{equation}
According to the Fermi's golden rule, neutrino emissivity in the
magnetic field can be written as
\begin{eqnarray}\label{402}
\epsilon_{B}=\frac{1}{VT}\sum_{n=0}^{\infty}\sum_{n'=0}^{\infty}\int
\frac{d^{2}p_{\y}L_{x}L_{z}}{(2\pi)^{2}}\frac{d^{3}kV}{(2\pi)^{3}}
\frac{d^{2}p'_{\y}L_{x}L_{z}}{(2\pi)^{2}}\frac{d^{3}k'V}{(2\pi)^{3}}
\frac{d^{3}q_{1}V}{(2\pi)^{3}}\frac{d^{3}q_{2}V}{(2\pi)^{3}}
F\left(\sum_{spins}|S|^{2}\right)\omega,
\end{eqnarray}
where $\omega$ is the total energy of neutrino with the momentum of
$q_{1}$ and antineutrino with the momentum of $q_{2}$, all other
momenta have the same definition as was given in the last section.
$F$ is defined in terms of the Fermi-Dirac distribution functions of
nucleons as
\begin{equation}\label{403}
F=f_{p}f_{n}(1-f_{p'})(1-f_{n'});\hspace{1.5cm}f=\frac{1}{1+\exp(\frac{E-\mu}{T})},
\end{equation}
in which $\mu$ is the chemical potential of the nucleon. Now we
consider the nucleons, comparing the available energies, as
non-relativistic particles.  Furthermore the anomalous magnetic
moment in the energy spectrum of nucleons are negligible, see Eqs.
(\ref{b12}), (\ref{b13}). Meanwhile since the neutrinos are soft the
three-momenta of neutrino and antineutrino can be dropped from the
Dirac $\delta$-functions.  Therefore the spin summed over the
squared scattering matrix element can be obtained as
\begin{eqnarray}\label{404}
\sum_{spin}|S|^{2}=\frac{L_{x}L_{z}T}
{(L_{x}L_{z})^{2}V^{4}2E_{1}2E_{2}}
(2\pi)^{3}\delta^{3}_{\y}(p'+k'+q_{1}+q_{2}-p-k)
\sum_{spin}|\mathcal{M}|^{2},
\end{eqnarray}
where
\begin{eqnarray}\label{405}
\sum_{spin}|\mathcal{M}|^{2}=\left(\frac{G_{F}g_{\pi
N}^{2}}{\sqrt{2}m_{N}^{2}m_{\pi}^{2}}\right)^{2}\frac{E_{1}E_{2}}{(E_{1}+E_{2})^{2}}
\vec{k}^{2}\{((p'_{z}-p_{z})^{2}+2eB(n'+n))\no\\
\times(I_{n,n'}I_{n,n'}^{\ast}+I_{n-1,n'-1}I_{n-1,n'-1}^{\ast})
-4eB\sqrt{n'n}(I_{n,n'}I_{n-1,n'-1}^{\ast}+I_{n-1,n'-1}I_{n,n'}^{\ast})\},
\end{eqnarray}
in which the functions $I_{n,m}$ are defined in (\ref{b2}).
Substituting (\ref{404}) in the emissivity expression (\ref{402})
yields
\begin{eqnarray}\label{406}
\epsilon_{B}=\sqrt{eB}\sum_{n=0}^{\infty}\sum_{n'=0}^{\infty}\int
\frac{d^{2}p_{\y}}{(2\pi)^{2}}\frac{d^{3}k}{(2\pi)^{3}}
\frac{d^{2}p'_{\y}}{(2\pi)^{2}}\frac{d^{3}k'}{(2\pi)^{3}}
\frac{d^{3}q_{1}}{(2\pi)^{3}}\frac{1}{2E_{1}}
\frac{d^{3}q_{2}}{(2\pi)^{3}}\frac{1}{2E_{2}}\no\\
\times(2\pi)^{3}\delta^{3}_{\y}(p'+k'+q_{1}+q_{2}-p-k)
F\left(\sum_{spins}|\mathcal{M}|^{2}\right)\omega.
\end{eqnarray}
One should note that the range of $L_{y}$ can be fixed in a natural
way as $L_{y}=\frac{1}{\sqrt{eB}}$ \cite{Liu}. Furthermore to
perform the integrals the momentum of particles for a degenerate
system, can not be larger than the Fermi momenta that in turns
depend on the density of system and implicitly on the temperature
\cite{friman}, so the phase space integrals of the initial particles
confined to the Fermi surface. While the remanning integrals on the
phase space of the final particles can be accomplished using the
momentum conserving Dirac $\delta$-functions .  Here we define
$\Upsilon=\frac{\epsilon_{B}}{\epsilon_{0}}$ where $\epsilon_{0}$
shows the corresponding emissivity in the absence of magnetic field.
The variation of $\Upsilon$ with respect to the neutron star
magnetic field for different temperature is shown in Fig.\ref{fig5}.

\section{Conclusion}
We have examined the nucleon-nucleon scattering in a large magnetic
background in a neutron star at the temperature about 1 \texttt{MeV}
. Since the energy is low we can safely use the one-pion-exchange
approximation and the four-fermion approximation. In fact the
external magnetic field can change the effective coupling constant
that is shown in the Fig.1. As the figure shows the coupling is
insensitive to the background field up to the $10^{17}\texttt{G}$
and then by increasing the magnetic field to $4\times
10^{18}\texttt{G}$ the squared value of the strong coupling constant
decreases to 0.6 its initial value. However the magnetic field can
change the nucleon wave functions and their corresponding field
operators as well, see equations (4) and (8). Therefore the
cross-section has an overall change as given in equation (34). The
variation of the cross-section, for head-on collisions, in a
magnetic background that is normalized with its correspondence
without background but with the same approximations are drawn in the
Fig.2 and Fig.3 for different temperatures. As the figures show the
deviation for the lower temperature start in the lower magnetic
fields. Nevertheless to have a significant deviation the magnetic
field even for the temperature as low as 0.1 \texttt{MeV}, should be
grater than $10^{15}\texttt{G}$.  We also studied the neutrino
bremsstrahlung emissivity by considering just one Feynman diagram
for this process at the tree level to give an estimate on the
neutrino emissivity in a strong magnetic field. As is shown in the
Fig.\ref{fig5} around $B= 10^{18}\texttt{G}$ the emissivity changes
about an order of magnitude with respect to its corresponding value
without magnetic filed.  Furthermore the  enhancement on the
emissivity is higher for the lower temperature. Therefore the
thermal evolution of the neutron stars in the late time cooling era
for $B\geq 10^{15}\texttt{G}$ can be dramatically changed.
\newpage
\appendix
\section{Solutions of Dirac equation}
In this Appendix we drive the wave functions of  proton and neutron
in the external magnetic field by tracing a procedure similar to
\cite{Mamsourov}. Dirac equation for a particle with mass $m$,
charge $e=|e|$ and anomalous magnetic moment
$\tilde{\mu}=K\frac{e}{2m}$ in presence of the external magnetic
field (\ref{101}) is
\begin{eqnarray}\label{a1}
i\frac{\partial\psi}{\partial t}&=&H_{D}\psi,\no\\
H_{D}&=&(\vec{\alpha}.\vec{\Pi}+\beta
m-\tilde{\mu}B\beta{\Sigma}_{z}).
\end{eqnarray}
Where the canonical momentum $\vec{\Pi}$ is given by the minimal
coupling
\begin{equation}\label{a2}
\vec{\Pi}=-i\vec{\nabla}-e\vec{A}.
\end{equation}
By using the standard Pauli-Dirac representation for Dirac
matrices
\begin{equation}\label{a3}
\vec{\alpha}=\left(%
\begin{array}{cc}
  0 & \vec{\sigma} \\
  \vec{\sigma} & 0 \\
\end{array}%
\right),
~~~~~\beta=\left(%
\begin{array}{cc}
  I & 0 \\
  0 & -I \\
\end{array}%
\right),
~~~~~\Sigma_{z}=\left(%
\begin{array}{cc}
  \sigma_{z} & 0 \\
  0 & \sigma_{z} \\
\end{array}%
\right),
~~~~~\gamma^{5}=\left(%
\begin{array}{cc}
  0 & I \\
  I & 0 \\
\end{array}%
\right),
\end{equation}
and general form of the wave function
\begin{equation}\label{a4}
\psi(t,\vec{x})=e^{-ip.x_{\y}}\left(%
\begin{array}{c}
u_{1}\\
u_{2}\\
u_{3}\\
u_{4}\\
\end{array}%
\right),
\end{equation}
the following system of equations will be obtained
\begin{eqnarray}\label{a5}
(E-m+\tilde{\mu}B)u_{1}-p_{z}u_{3}-(\Pi_{x}-i\Pi_{y})u_{4}&=&0,\no\\
(E-m-\tilde{\mu}B)u_{2}+p_{z}u_{4}-(\Pi_{x}+i\Pi_{y})u_{3}&=&0,\no\\
(E+m-\tilde{\mu}B)u_{3}-p_{z}u_{1}-(\Pi_{x}-i\Pi_{y})u_{2}&=&0,\no\\
(E+m+\tilde{\mu}B)u_{4}+p_{z}u_{2}-(\Pi_{x}+i\Pi_{y})u_{1}&=&0.
\end{eqnarray}
To solve equations (\ref{a5}) one can assume
\begin{eqnarray}\label{a6}
\!\!\!\!u_{1}=C_{1}I_{n}(\xi),~~~~~u_{2}=C_{2}I_{n-1}(\xi),~~~~~u_{3}=C_{3}I_{n}(\xi),
~~~~~u_{4}=C_{4}I_{n-1}(\xi),
\end{eqnarray}
where the function $I_{n}(\xi)$ is defined as
\begin{eqnarray}\label{a7}
I_{n}(\xi)&\equiv& N_{n}e^{-\xi^{2}/2}H_{n}(\xi);~~~~~n=0,1,2,\ldots\no\\
\xi&\equiv&\sqrt{eB}y+\frac{p_{x}}{\sqrt{eB}},~~~~~N_{n}=\sqrt{\frac{\sqrt{eB}}{\sqrt{\pi}2^{n}n!}}.
\end{eqnarray}
The function $H_{n}(\xi)$ is the Hermit polynomial and in according
to the definition $H_{-1}(\xi)=0$. Using this choice for normalization
constant $N_{n}$, it can be showed that the functions $I_{n}(\xi)$
satisfy the completeness relation
\begin{equation}\label{a8}
\sum_{n=0}^{\infty}I_{n}(\xi)I_{n}(\xi')=\delta(y-y'),
\end{equation}
where $\xi'$ is given by replacing $y$ by $y'$ in the definition
of the $\xi$ (\ref{a7}). After substituting equations (\ref{a6})
in the system of equations (\ref{a5}), the system of linear equations
for coefficients $C_{i}$ will be obtained as follows
\begin{eqnarray}\label{a9}
(E\mp m\pm\tilde{\mu}B)C_{1,3}-p_{z}C_{3,1}-\sqrt{2neB}C_{4,2}&=0&,\no\\
(E\mp
m\mp\tilde{\mu}B)C_{2,4}+p_{z}C_{4,2}-\sqrt{2neB}C_{3,1}&=0&.
\end{eqnarray}
Equations (\ref{a9}) have a non trivial solutions for $C_{i}$ if
the Determinant of the coefficients be zero. Imposing this
condition yields the spectrum of the energy
\begin{equation}\label{a10}
E_{n,s}=\sqrt{p_{z}^{2}+(\sqrt{m^{2}+2neB}-s\tilde{\mu}B)^{2}};~~~~~s=\pm1.
\end{equation}
In the energy spectrum (\ref{a10}) index of $n$ is called Landau
level. To solve equations (\ref{a9}) it is better to introduce the
fermion spin projection operator on the direction of the magnetic
field as
\begin{equation}\label{a11}
\hat{\mu}_{z}=m\Sigma_{z}+i\gamma^{5}\beta(\vec{\Sigma}\times\vec{\Pi})_{z}.
\end{equation}
Since the operator $\hat{\mu}_{z}$ commutes with the
Dirac Hamiltonian $H_{D}$, the wave functions are eigenstate of this
operator
\begin{equation}\label{a12}
\hat{\mu}_{z}\psi=\mu_{0}\psi,
\end{equation}
where $\mu_{0}$ is the eigenvalue of the operator $\hat{\mu}_{z}$.
Then using the equations (\ref{a12}) and (\ref{a6}), one leads to
the following system of liner equations for the coefficients
$C_{i}$
\begin{eqnarray}\label{a13}
\!\!\!\!\!\!(m\mp\mu_{0})C_{1,4}\pm\sqrt{2neB}C_{4,1}=0,
~~~~~~~~~~~~(m\pm\mu_{0})C_{2,3}\pm\sqrt{2neB}C_{3,2}=0.
\end{eqnarray}
Again by imposing the condition that, Determinant of the
coefficients must be zero, $\mu_{0}$ can be determined as
\begin{equation}\label{a14}
\mu_{0}=s\sqrt{m^{2}+2neB};~~~~~s=\pm1.
\end{equation}
Now by using the equations (\ref{a9}), (\ref{a13}) and
the following normalization condition, the coefficients $C_{i}$ can be
determined
\begin{equation}\label{a15}
\sum_{i=1}^{i=4}|C_{i}|^{2}=1.
\end{equation}
Therefore the wave function can be obtained as
\begin{eqnarray}\label{a16}
\psi_{s}(x)&=&N_{s}e^{-ip.x_{\y}}U_{s}(y,n,\vec{p}_{\y});\no\\
U_{s}(y,n,\vec{p}_{\y})&=&\left(%
\begin{array}{c}
(\frac{1+s}{2}+ \frac{1-s}{2}\tilde{p}_{zs}\tilde{\gamma}_{n})I_{n}(\xi)\\
(-\frac{1+s}{2}\tilde{p}_{zs}\tilde{\gamma}_{n}+\frac{1-s}{2})I_{n-1}(\xi)\\
(\frac{1+s}{2}\tilde{p}_{zs}+ \frac{1-s}{2}\tilde{\gamma}_{n})I_{n}(\xi) \\
(\frac{1+s}{2}\tilde{\gamma}_{n}- \frac{1-s}{2}\tilde{p}_{zs})I_{n-1}(\xi) \\
\end{array}%
\right),~~~~~~s=\pm1,
\end{eqnarray}
where the quantities $\tilde{p}_{zs}$ and $\tilde{\gamma}_{n}$ are
defined as
\begin{eqnarray}\label{a17}
\!\!\!\!\!\!\tilde{p}_{zs}=\frac{p_{z}}{E_{n,s}+\sqrt{E_{n,s}^{2}-p_{z}^{2}}},
~~~~~~~~~~~~\tilde{\gamma}_{n}=\frac{\sqrt{2neB}}{m+\sqrt{m^{2}+2neB}},
\end{eqnarray}
and the normalization constant $N_{s}$ of the wave function is
\begin{equation}\label{a18}
N_{s}=\frac{1}{\sqrt{(1+\tilde{p}_{zs}^{2})(1+\tilde{\gamma}_{n}^{2})}}.
\end{equation}
One can easily find the wave functions for the negative energy spinors by doing the
similar procedure. In this case it is easier to determine the upper
component with respect to the lower one. After some manipulations one has
\begin{eqnarray}\label{a19}
\psi_{s}(x)&=&N_{s}e^{ip.x_{\y}}V_{s}(y,n,\vec{p}_{\y});\no\\
\!\!\!\!\!\!\!\!\!\!\!\!\!\!\!V_{s}(y,n,\vec{p}_{\y})&=&\left(%
\begin{array}{c}
(\frac{1+s}{2}\tilde{p}_{zs}-\frac{1-s}{2}\tilde{\gamma}_{n})I_{n}(\tilde{\xi})\\
(-\frac{1+s}{2}\tilde{\gamma}_{n}-\frac{1-s}{2}\tilde{p}_{zs})I_{n-1}(\tilde{\xi})\\
(\frac{1+s}{2}-\frac{1-s}{2}\tilde{p}_{zs}\tilde{\gamma}_{n})I_{n}(\tilde{\xi})\\
(\frac{1+s}{2}\tilde{p}_{zs}\tilde{\gamma}_{n}+\frac{1-s}{2})I_{n-1}(\tilde{\xi})\\
\end{array}%
\right);~~~~~~\tilde{\xi}\equiv\sqrt{eB}y-\frac{p_{x}}{\sqrt{eB}},
\end{eqnarray}
where quantities $\tilde{p}_{zs}$ and $\tilde{\gamma}_{n}$ and
normalization constant $N_{s}$ have the same definition as the
positive energy solutions and can be determined, respectively,
from equations (\ref{a17}) and (\ref{a18}). Neutron has not charge
but due to its anomalous magnetic moment its wave function and
energy spectrum can be modified by the magnetic field. Solving
Dirac equation for this case is simpler than proton case. By
doing the same procedure as the proton case, the spectrum can be
obtained as
\begin{equation}\label{a20}
E_{s}=\sqrt{p_{z}^{2}+(\sqrt{m^{2}+p_{\bot}^{2}}-s\tilde{\mu}B)^{2}};~~~~~s=\pm1,
\end{equation}
and the wave functions will be given by
\begin{eqnarray}\label{a21}
\psi_{s}(x)&=&N_{s}e^{-ip.x}U_{s}(\vec{p});\no\\
U_{s}(\vec{p})&=&\left(%
\begin{array}{c}
(\frac{1+s}{2}+ \frac{1-s}{2}\tilde{p}_{zs}\tilde{p}_{s})\\
(-\frac{1+s}{2}\tilde{p}_{zs}\tilde{p}_{s}+\frac{1-s}{2})\\
(\frac{1+s}{2}\tilde{p}_{zs}+\frac{1-s}{2}\tilde{p}_{s})\\
(\frac{1+s}{2}\tilde{p}_{s}-\frac{1-s}{2}\tilde{p}_{zs})\\
\end{array}%
\right),~~~~~s=\pm1,
\end{eqnarray}
where quantities $\tilde{p}_{zs}$, $\tilde{p}_{s}$ and $p_{\bot}$
are defined by
\begin{eqnarray}\label{a22}
\!\!\!\!\!\!\!\tilde{p}_{zs}&=&\frac{p_{z}}{E_{s}+\sqrt{E_{s}^{2}-p_{z}^{2}}},
~~~~~~~~~~~~~~\tilde{p}_{s}=\frac{p_{s}}{m+\sqrt{m^{2}+p_{\bot}^{2}}};\no\\
p_{s}&=&p_{x}+sip_{y},~~~~~~~~~~~~~~~~~~~~~~p_{\bot}=\sqrt{p_{+}p_{-}},
\end{eqnarray}
and the normalization constant given by
\begin{equation}\label{a23}
\!\!\!\!\!\!\!\!\!\!\!\!\!\!\!\!N_{s}=\frac{1}{\sqrt{(1+\tilde{p}_{zs}^{2})(1+\tilde{p}_{\bot}^{2})}};
~~~~~~~~~~~~\tilde{p}_{\bot}=\sqrt{\tilde{p}_{+}\tilde{p}_{-}}.
\end{equation}
Finally the solutions for the negative energy case can be obtained as
\begin{eqnarray}\label{a24}
\psi_{s}(x)&=&N_{s}e^{ip.x}V_{s}(\vec{p});\no\\
V_{s}(\vec{p})&=&\left(%
\begin{array}{c}
(\frac{1+s}{2}\tilde{p}_{zs}+ \frac{1-s}{2}\tilde{p}_{s})\\
(\frac{1+s}{2}\tilde{p}_{s}-\frac{1-s}{2}\tilde{p}_{zs})\\
(\frac{1+s}{2}+\frac{1-s}{2}\tilde{p}_{zs}\tilde{p}_{s})\\
(-\frac{1+s}{2}\tilde{p}_{zs}\tilde{p}_{s}+\frac{1-s}{2})\\
\end{array}%
\right).
\end{eqnarray}
Definition of the parameters in the wave functions (\ref{a24}) are
the same as the equations (\ref{a22}) and (\ref{a23}).
\section{Average of the squared scattering matrix elements}
The matrix elements that are sum of the neutral
(\ref{217}) and the charge current amplitude (\ref{218}), contain the following integrals
\begin{eqnarray}\label{b1}
I_{n,m}=\int{dy}e^{-i(k'_{y}-k_{y})y}I_{n}(\xi)I_{m}(\xi_{\ast}),
\end{eqnarray}
where $\xi_{\ast}$ is given by replacing $p_{x}$ by $p'_{x}$ in
the definition of the $\xi$ (\ref{a7}). Details of this
integration can be found in \cite{duane}, however the final answer is as
follows
\begin{eqnarray}\label{b2}
I_{n,m}&=&\sqrt{\frac{n!}{m!}}e^{-\eta^{2}/2}e^{i\phi_{0}}
(\eta_{x}+i\eta_{y})^{^{m-n}}L_{n}^{m-n}(\eta^{2});~~~~~~~m\geq
n\geq0,\no\\
I_{n,m}&=&\sqrt{\frac{m!}{n!}}e^{-\eta^{2}/2}e^{i\phi_{0}}
(-\eta_{x}+i\eta_{y})^{^{n-m}}L_{m}^{n-m}(\eta^{2});~~~~~~~n\geq
m\geq0,
\end{eqnarray}
in which  $L_{n}^{k}$ is the generalized Laguerre polynomial and
variables $\eta$ and $\phi_{0}$ are defined as
\begin{eqnarray}\label{b3}
\eta_{x}&=&\frac{p'_{x}-p_{x}}{\sqrt{2eB}},~~~~~~~~\eta_{y}=-\frac{k'_{y}-k_{y}}{\sqrt{2eB}},\no\\
\eta^{2}&=&\eta_{x}^{2}+\eta_{y}^{2},~~~~~~~~\phi_{0}=\frac{(p'_{x}+p_{x})(k'_{y}-k_{y})}{2eB}.
\end{eqnarray}
By using the explicit form of the wave functions and the definition
(\ref{b1}), one can find the nonrelativistic leading order
term in the amplitude as follows
\begin{equation}\label{b9}
\mathcal{M}=\mathcal{M}_{nc}+\mathcal{M}_{cc},
\end{equation}
such that the neutral current amplitude matrix element is given by
\begin{eqnarray}\label{b10}
\mathcal{M}_{nc}&=&\frac{g_{\pi
N}^{2}}{m_{\pi}^{2}}\Big[(\frac{1+r}{2})(\frac{1+r'}{2})(\tilde{k}'_{z}-\tilde{k}_{z})+
(\frac{1+r}{2})(\frac{1-r'}{2})(\tilde{k}'_{+}-\tilde{k}_{+})\no\\
&+&(\frac{1-r}{2})(\frac{1+r'}{2})(\tilde{k}'_{-}-\tilde{k}_{-})-
(\frac{1-r}{2})(\frac{1-r'}{2})(\tilde{k}'_{z}-\tilde{k}_{z})
\Big]\no\\
&\times&\Big[\{(\frac{1+s}{2})(\frac{1+s'}{2})(\tilde{p}'_{z}-\tilde{p}_{z})
+(\frac{1+s}{2})(\frac{1-s'}{2})\tilde{\gamma}_{n'}
-(\frac{1-s}{2})(\frac{1+s'}{2})\tilde{\gamma}_{n}\}I_{n,n'}\no\\
&-&\{(\frac{1-s}{2})(\frac{1-s'}{2})(\tilde{p}'_{z}-\tilde{p}_{z})
-(\frac{1-s}{2})(\frac{1+s'}{2})\tilde{\gamma}_{n'}
+(\frac{1+s}{2})(\frac{1-s'}{2})\tilde{\gamma}_{n}\}I_{n-1,n'-1}\Big],\no\\
\end{eqnarray}
and the charged current amplitude is given by
\begin{eqnarray}\label{b11}
\mathcal{M}_{cc}&=&\frac{2Dg_{\pi N}^{2}}{m_{\pi}^{2}}\Big[
\{(\frac{1+r}{2})(\frac{1+s'}{2})(\tilde{p}'_{z}-\tilde{k}_{z})
+(\frac{1+r}{2})(\frac{1-s'}{2})\tilde{\gamma}_{n'}
-(\frac{1-r}{2})(\frac{1+s'}{2})\tilde{k}_{-}\}\no\\
&\times&\{(\frac{1+s}{2})(\frac{1+r'}{2})(\tilde{k}'_{z}-\tilde{p}_{z})
+(\frac{1+s}{2})(\frac{1-r'}{2})\tilde{k'}_{+}
-(\frac{1-s}{2})(\frac{1+r'}{2})\tilde{\gamma}_{n}\}I_{n,n'}\no\\
&+&\{-(\frac{1+r}{2})(\frac{1-s'}{2})\tilde{k}_{+}
+(\frac{1-r}{2})(\frac{1+s'}{2})\tilde{\gamma}_{n'}
-(\frac{1-r}{2})(\frac{1-s'}{2})(\tilde{p}'_{z}-\tilde{k}_{z})\}\no\\
&\times&\{(\frac{1+s}{2})(\frac{1+r'}{2})(\tilde{k}'_{z}-\tilde{p}_{z})
+(\frac{1+s}{2})(\frac{1-r'}{2})\tilde{k'}_{+}
-(\frac{1-s}{2})(\frac{1+r'}{2})\tilde{\gamma}_{n}\}I_{n,n'-1}\no\\
&+&\{(\frac{1+r}{2})(\frac{1+s'}{2})(\tilde{p}'_{z}-\tilde{k}_{z})
+(\frac{1+r}{2})(\frac{1-s'}{2})\tilde{\gamma}_{n'}
-(\frac{1-r}{2})(\frac{1+s'}{2})\tilde{k}_{-}\}\no\\
&\times&\{-(\frac{1+s}{2})(\frac{1-r'}{2})\tilde{\gamma}_{n}
+(\frac{1-s}{2})(\frac{1+r'}{2})\tilde{k'}_{-}
-(\frac{1-s}{2})(\frac{1-r'}{2})(\tilde{k}'_{z}-\tilde{p}_{z})\}I_{n-1,n'}\no\\
&+&\{(\frac{1+r}{2})(\frac{1-s'}{2})\tilde{k}_{+}
-(\frac{1-r}{2})(\frac{1+s'}{2})\tilde{\gamma}_{n'}
+(\frac{1-r}{2})(\frac{1-s'}{2})(\tilde{p}'_{z}-\tilde{k}_{z})\}\no\\
&\times&\{(\frac{1+s}{2})(\frac{1-r'}{2})\tilde{\gamma}_{n}
-(\frac{1-s}{2})(\frac{1+r'}{2})\tilde{k'}_{-}
+(\frac{1-s}{2})(\frac{1-r'}{2})(\tilde{k}'_{z}-\tilde{p}_{z})\}I_{n-1,n'-1}\Big].\no\\
\end{eqnarray}
Using the following nonrelativistic expansion of the energy for proton as
\begin{equation}\label{b12}
E_{n,s}=m_{N}+\frac{p_{z}^{2}}{2m_{N}}+\frac{neB}{m_{N}}-s\tilde{\mu}_{p}B,
\end{equation}
and for the neutron as
\begin{equation}\label{b13}
E_{s}=m_{N}+\frac{\vec{p}^{2}}{2m_{N}}-s\tilde{\mu}_{n}B,
\end{equation}
along with the equations (\ref{a17}) and (\ref{a22}) lead to
\begin{eqnarray}\label{b14}
\tilde{p}_{z}=\frac{p_{z}}{2m_{N}},
~~~~~~~~\tilde{\gamma}_{n}=\frac{\sqrt{2neB}}{2m_{N}},
~~~~~~~~\tilde{k}_{z}=\frac{k_{z}}{2m_{N}},
~~~~~~~~\tilde{k}_{\pm}=\frac{k_{\pm}}{2m_{N}},
\end{eqnarray}
with the similar definitions for the primed quantities.


\newpage
\begin{thebibliography}{99}
\bibitem{haberl}
F.~Haberl, Astrophys. and Space Sci. J. {\bf 308}, 181 (2007).
\bibitem{tsuruta}
S.~Tsuruta, Phys. Rep. {\bf292}, 1 (1998).
\bibitem{thompson}
C.~Thompson and R.~C.~Duncan,
Astrophys.\ J.\ {\bf 408}, 194 (1993).
\bibitem{blandford}
R.~D.~Blandford, J.~H.~Applegate and L.~Hernquist, Mon. Not. Roy. Astr. Soc. {\bf 204}, 1025 (1983).
\bibitem{Duncan}
R.~C.~Duncan and C.~Thompson,
Astrophys.\ J.\  {\bf 392}, L9 (1992);
C.~Thompson and R.~C.~Duncan,
Astrophys.\ J.\  {\bf 473}, 322 (1996);
C.~Kouveliotou {\it et al.},
Nature {\bf 393}, 235 (1998).
\bibitem{Mereghetti}
S.~Mereghetti,
Astron.\ Astrophys.\ Rev.\  {\bf 15}, 225 (2008)
[arXiv:0804.0250 [astro-ph]].
\bibitem{manchester}
R.~N.~Manchester and J.~H.~Taylor, Astron. J.{\bf 86}, 1953 (1981).
\bibitem{max}
D.~Lai,
Rev.\ Mod.\ Phys.\ {\bf 73}, 629 (2001) [arXiv:astro-ph/0009333].
\bibitem{itoh}
N.~Itoh, Mon. Not. Roy. Astr. Soc. {\bf 173}, 1 (1975).
\bibitem{aguilera}
Deborah N.~Aguilera, Jose A.~Pons and Juan A.~Miralles, Astron. Astrophys. J. {\bf 486}, 255 (2008).
\bibitem{hanhart}
C.~Hanhart, D.~R.~Phillips and S.~ Reddy, Phys. Lett. B {\bf 499}, 9 (2001);
C.~Hanhart, D.~R.~Phillips, S.~ Reddy and M.~J.~Savage, Nucl. Phys. B {\bf 595}, 335 (2001).
\bibitem{friman}
B.~L.~Friman and O.~V.~Maxwell, Astrophys.\ J.\  {\bf 232}, 541 (1979).
\bibitem{Erdas}
A.~Erdas and G.~Feldman,
Nucl.\ Phys.\  B {\bf 343}, 597 (1990).
\bibitem{Hannestad}
W.~Keil, H.~T.~Janka and G.~Raffelt,
Phys.\ Rev.\  D {\bf 51}, 6635 (1995) [arXiv:hep-ph/9410229];
S.~Hannestad and G.~Raffelt,
Astrophys.\ J.\  {\bf 507}, 339 (1998) [arXiv:astro-ph/9711132].
\bibitem{Liu}
X.~W.~Liu, X.~P.~Zheng and D.~F.~Hou,
Astropart.\ Phys.\ {\bf 24}, 92 (2005) [arXiv:astro-ph/0412515].
\bibitem{neu-emmission}
Mario Riquelme, Andreas Reisenegger, Olivier Espinosa and Claudio O.
Dib, Astrophys.\ J.\ {\bf 439}, 427 (2005).
\bibitem{neu-absor}
Huaiyu Duan and Yong-Zhong Qian, Phys.\ Rev.\ D {\bf 72}, 023005
(2005).
\bibitem{Mamsourov}
I.~Mamsourov and H.~Goudarzi,
[arXiv:hep-ph/0404086].
\bibitem{duane}
D.~A.~Dicus, W.~W.~Repko and T.~M.~Tinsley,
Phys.\ Rev.\ D {\bf 76}, 025005 (2007) [Erratum-ibid.\ D {\bf 76}, 089903 (2007)]
[arXiv:0704.1695 [hep-ph]].
\end{thebibliography}
\end{document}